\documentclass[draft]{agujournal2019}
\usepackage{url}
\usepackage{lineno}
\usepackage[inline]{trackchanges}
\usepackage{amsmath}
\usepackage{graphicx}
\usepackage[letterpaper,margin=1in]{geometry}

\draftfalse   

\journalname{Geophysical Research Letters}

\newcommand{\NDM}{NDM}
\newcommand{\msis}{NRLMSISE-00}
\newcommand{\kp}{\ensuremath{K_p}}

\begin{document}

\title{The Neural Density Model: An Open 3D Thermospheric Density Reanalysis for 2003--2026 From Tracked Space Debris}

\authors{Enrico Camporeale\affil{1,2}}

\affiliation{1}{School of Physical and Chemical Sciences, Queen Mary University of London, London, UK}
\affiliation{2}{Space Weather Technology, Research and Education Center (SWx TREC), University of Colorado Boulder, Boulder, CO, USA}

\correspondingauthor{Enrico Camporeale}{enrico.camporeale@qmul.ac.uk}

\begin{keypoints}
\item We present NDM, an open 3D thermospheric density reanalysis for 2003 to 2026 at 250 to 800 km, built only from public observations
\item A differentiable drag operator fits a neural density correction directly to the orbital decay of 600 to 800 tracked objects per year
\item On withheld satellites the error is 23 percent below HASDM and fourfold below NRLMSISE-00; on independent Spire orbits it matches HASDM
\end{keypoints}

\begin{abstract}
Thermospheric mass density controls satellite drag but is observed directly by few missions, and the leading operational product (HASDM) relies on non-public radar data. We present the Neural Density Model (NDM), a 3D density reanalysis for October 2003--October 2026 at 250--800~km. Previous approaches either convert the orbital decay of tracked objects into point densities or correct empirical models with a few global scale factors. NDM instead fits a spatially resolved neural correction to NRLMSISE-00 directly to the observed decay of 600--800 objects per year, through a differentiable drag operator, jointly with their ballistic coefficients, calibration spheres and accelerometer densities. On withheld satellites the median error is 0.078 (natural-log units), against 0.101 for HASDM and 0.310 for NRLMSISE-00; NDM has the smaller bias, HASDM slightly smaller scatter. On Spire GNSS orbits, never used in training, NDM and HASDM perform comparably. NDM uses only public observations.
\end{abstract}

\section*{Plain Language Summary}
Satellites in low Earth orbit are slowed by the thin upper atmosphere, and how strongly depends on its density, which can double within hours during geomagnetic storms. Predicting satellite positions, avoiding collisions and planning re-entries all depend on knowing this density, but only a handful of research satellites have ever measured it directly. We turn the space debris and spent rocket stages already tracked by radar into a global network of density sensors: as they lose altitude, the rate of their decay reveals the density they have flown through. A machine-learning model, trained jointly on the decay of 600--800 such objects per year and on the few direct measurements, reconstructs the density between 250 and 800~km every day from 2003 to 2026. Tested against satellites it never saw, it is about as accurate as the best operational model, which is built on restricted military data, and four times more accurate than the standard empirical model. This open record can serve as a reference for satellite operators and as a training set for machine-learning forecasts of space weather.

\section{Introduction}

Atmospheric drag is the largest and least predictable non-gravitational force on objects in low Earth orbit (LEO), and its uncertainty is dominated by the uncertainty in thermospheric mass density \cite{Emmert2015,Berger2020}. With more than ten thousand active satellites and a growing debris population, density errors propagate directly into conjunction assessment, re-entry prediction and constellation operations. The February 2022 loss of 38 Starlink satellites during a moderate storm \cite{Fang2022,Berger2023} and the large drag-induced orbit changes during the May 2024 superstorm \cite{Parker2024} illustrate the stakes.

Empirical models such as \msis\ \cite{Picone2002} and NRLMSIS~2.1 \cite{Emmert2022} have systematic errors of tens of percent, reaching a factor of about two at solar minimum, and their storm-time errors are larger still \cite{Bruinsma2018,Oliveira2019,Emmert2015}. Direct observations are sparse: accelerometer and precise-orbit densities from CHAMP, GRACE, GOCE, Swarm and GRACE-FO \cite{Doornbos2010,Siemes2016,vandenIJssel2020,Siemes2023} sample only a few orbital planes and altitudes at any time. Data-assimilation studies have corrected empirical or physics-based models using accelerometer densities \cite{Forootan2021,Forootan2023,Ponder2023,Cheng2024,Dietrich2026}, GNSS orbits of CubeSats \cite{Sutton2021}, or drag of a few objects \cite{Brandt2020,Mutschler2026}, and machine learning has been used to emulate, calibrate or forecast density \cite{Mehta2017,Licata2022,Huang2026,Wang2026}. Most of these efforts adjust a handful of global drivers or scale factors of an existing model over days to a year. The spatial pattern of the density, that is, how it varies with latitude, local time and altitude, therefore remains that of the underlying model. Reduced-dynamic and energy-dissipation retrievals from single satellites \cite{Ray2024,Constant2026} are accurate along their own orbits but are not global.

The orbital decay of the objects routinely tracked by the US Space Surveillance Network, distributed as two-line element sets (TLEs), is a free, continuous and global record of drag \cite{Picone2005,Emmert2008,Doornbos2008}. The operational state of the art, the High Accuracy Satellite Drag Model \cite<HASDM;>{Storz2005}, also uses the observed orbital decay of tracked objects to correct an empirical model. It dynamically calibrates a Jacchia-type empirical model \cite<originally Jacchia 1970, JB2008 in current versions;>{Jacchia1970,Bowman2008} using radar observations of several tens of calibration objects. HASDM densities are available to researchers on request \cite{Tobiska2021}, but the radar observations, the calibration-object set and the processing are not public, so the product cannot be independently reproduced, extended or audited.

Weather forecasting was transformed by reanalyses such as ERA5 \cite{Hersbach2020}: long, gap-free, observation-constrained reconstructions of the atmospheric state that serve both as a climatological reference and as training data for machine-learning forecast models. No equivalent open product exists for the thermosphere. Here we present the Neural Density Model (\NDM), a 23-year 3D density reanalysis built entirely from public or openly licensed data. Three elements make it possible: (1) a neural parameterization of the density field whose parameters are fitted by gradient descent through a fully differentiable drag observation operator, so that the orbital decay of each object constrains the density field directly, without first being converted into point estimates of density; (2) training on the orbital decay of 600--800 rocket bodies and debris objects per year (about $2.8\times10^6$ drag intervals over 2003--2026), anchored in absolute scale by spheres of known area-to-mass ratio and by accelerometer densities; and (3) a validation design that tests the product only on data it has not seen, including the GNSS orbits of the Spire constellation, which never enter training, and that separates the contribution of each data source.

\section{Method}

\subsection{Neural Parameterization of the Density Field}

\NDM\ represents the log-density as a correction to \msis:
\begin{equation}
\ln \rho(\mathbf{x},t) = \ln \rho_{\mathrm{MSIS}}(\mathbf{x},t) + \delta(\mathbf{x},t), \qquad
\delta = c_d + f_{\theta}(\mathbf{x}, \mathbf{z}_d),
\label{eq:model}
\end{equation}
where $d$ is the day containing $t$. The input vector $\mathbf{x}$ contains latitude, local solar time and day of year (as sines and cosines), altitude, daily and 81-day F10.7, \kp, and the current and lagged 3-hourly $a_p$ (previous 9~h, 12--33~h and 36--57~h). The network $f_\theta$ is a multilayer perceptron (3 hidden layers of 128 units, SiLU activations). The daily level $c_d$ and a four-dimensional latent vector $\mathbf{z}_d$ are free parameters fitted to that day's observations; they carry the day-to-day state that the indices do not explain, while $f_\theta$ encodes how that state maps onto latitude, local time and altitude. To keep the daily state from jumping between days, the loss includes a random-walk penalty
\begin{equation}
\mathcal{R} = \lambda_s \sum_d \left[ (c_{d+1}-c_d)^2 + \lVert \mathbf{z}_{d+1}-\mathbf{z}_d \rVert^2 \right],
\label{eq:smooth}
\end{equation}
which penalizes day-to-day changes unless the data require them; days with few observations then borrow information from their neighbors. The weight $\lambda_s$ is selected for each year from \{1, 10, 100\} on validation weeks. \msis\ itself is evaluated with its full geomagnetic input, the 3-hourly $a_p$ history of the preceding 57~h (its storm-time option), rather than with the daily $A_p$ alone. One network is trained per calendar year (with $\pm$15-day overlaps) as an ensemble of three seeds; the yearly fields are concatenated into the reanalysis.

The observations therefore constrain the field on time scales of one day and longer, through $(c_d,\mathbf{z}_d)$ and the 1--60-day TLE intervals. Variability on shorter time scales enters only through the geomagnetic indices in $\mathbf{x}$ and through \msis: it is model-driven, with a response learned from the observations, not observed directly.

\subsection{A Differentiable Drag Observation Operator}

Each TLE gives the mean motion $n$ of an object, the average angular rate of its orbit (for example in radians per second), which is related to the semi-major axis $a$ by $n^2a^3=\mu$, where $\mu$ is Earth's gravitational parameter. Drag removes orbital energy, so the orbit shrinks and $n$ increases. For an object $i$ with consecutive TLEs at times $t_1$ and $t_2$, the observation is therefore the change in mean motion, $\Delta n_i^{\mathrm{obs}} = n(t_2)-n(t_1)$, computed from the two SGP4 mean elements \cite{Vallado2006}; it is a change in orbital angular rate, not in density. Following the energy-dissipation approach of \citeA{Picone2005}, this change is predicted by integrating the drag power along the SGP4 trajectory, written as a sum over sample points $k$ between $t_1$ and $t_2$:
\begin{equation}
\Delta n_i^{\mathrm{pred}} = B_i \sum_k w_k\, \rho(\mathbf{x}_k, t_k), \qquad
w_k = \frac{3\, a\, n}{2\mu}\, |\mathbf{v}_{\mathrm{rel}}|\,(\mathbf{v}_{\mathrm{rel}}\cdot\mathbf{v})\,\Delta t_k ,
\label{eq:drag}
\end{equation}
where $\mathbf{v}$ is the orbital velocity, $\mathbf{v}_{\mathrm{rel}}$ the velocity relative to a co-rotating atmosphere, $\Delta t_k$ the sample spacing, and $B_i = C_D A/m$ the ballistic coefficient. The weights $w_k$ depend only on the trajectory and are computed once in advance. The density at each sample point is given by the model of Equation~\ref{eq:model}, so that, substituting Equation~\ref{eq:model} into Equation~\ref{eq:drag},
\begin{equation}
\Delta n_i^{\mathrm{pred}} = B_i \sum_k w_k\, \rho_{\mathrm{MSIS}}(\mathbf{x}_k,t_k)\, \exp\!\left[ c_{d(k)} + f_{\theta}\big(\mathbf{x}_k, \mathbf{z}_{d(k)}\big) \right],
\label{eq:chain}
\end{equation}
where $d(k)$ is the day of sample $k$. Equation~\ref{eq:chain} is an explicit, differentiable function of every unknown: the network weights $\theta$, the daily states $(c_d,\mathbf{z}_d)$ and the ballistic coefficients $B_i$. All of them are therefore estimated together by minimizing a single loss,
\begin{equation}
\mathcal{L} = w_{\mathrm{TLE}} \sum_{i} \frac{\kappa_i}{\sigma_i^2} \sum_j h\!\left(\Delta n_{ij}^{\mathrm{obs}} - \Delta n_{ij}^{\mathrm{pred}}\right) + \sum_{m} h\!\left( \ln \frac{\rho_m^{\mathrm{obs}}}{\rho_{\mathrm{MSIS},m}} - \delta_m \right) + \mathcal{R},
\label{eq:loss}
\end{equation}
where $j$ runs over the TLE intervals of object $i$, $m$ over accelerometer density samples, $h$ is the robust Huber function, $\sigma_i$ is the noise level of object $i$ (estimated from its residual scatter), $\kappa_i$ is a weight for the calibration objects (Section~2.3), and $w_{\mathrm{TLE}}=0.3$ balances the two data types. The gradients of $\mathcal{L}$ are obtained by automatic differentiation \cite<PyTorch;>{Paszke2019} and the minimization uses the Adam optimizer \cite{Kingma2015}. Differentiable propagation has previously been demonstrated on simulated orbits \cite{Izzo2024,Acciarini2024}; here it is applied to real tracking data at scale.

Because SGP4 mean elements remove the main zonal-harmonic effects, the remaining non-drag contributions to $\Delta n^{\mathrm{obs}}$ are small for the selected objects; they are treated as noise by the per-object error model and screening described below. We checked Equation~\ref{eq:drag} against full numerical integration (DOP853, J2--J6 gravity plus drag) for 114 intervals stratified over 32 classes of perigee altitude (285--2000~km), eccentricity and object type, including intervals with \kp~$\geq$~6. The ratio of predicted to integrated $\Delta n$ is $1.027$ (10th--90th percentile: 1.015--1.039), equal within 0.4\% for storm and quiet intervals and across perigee, eccentricity and object type; the constant offset is absorbed by the fitted $B_i$ (and biases the fixed-$B$ spheres, for which the ratio is 1.012--1.013, by about 1.2\%). On real data, the TLEs of the accelerometer satellites themselves (1,443 intervals) agree with TU Delft densities integrated along the same samples to 5--7\% per interval, with no dependence on geomagnetic activity or interval length.

\subsection{Observations}

\textit{TLE drag.} Each year we select up to 400 rocket bodies and 400 debris objects from the Space-Track catalog, stratified by perigee between 250 and 800~km. For each object, consecutive TLE pairs (at least one day apart) are merged into longer intervals until the drag signal is clearly larger than the TLE noise. Specifically, we require that the change in mean motion expected from \msis\ over the interval be at least five times the noise of a difference of two TLE mean motions, the latter estimated from the scatter of successive TLEs of that object. If more than 2\% of an object's intervals still show no decay ($\Delta n\leq0$), the target length is doubled, up to 60 days. Intervals are therefore longer when drag is weak, at high altitude and low solar activity. Each object has one free $B_i$ per year with no prior; its drag coefficient is computed per sample from a physical gas--surface model \cite<diffuse reflection with SESAM accommodation;>{Sentman1961,Pilinski2013}, so that changes of $C_D$ with altitude and composition are modeled and $B_i$ effectively estimates a time-averaged area-to-mass ratio. For tumbling debris this average is taken over intervals much longer than the tumbling periods. Residual changes of $B_i$ within a year appear as excess scatter, which raises $\sigma_i$ and lowers the weight of that object in Equation~\ref{eq:loss}.

\begin{figure}[ht]
\centering
\includegraphics[width=0.78\textwidth]{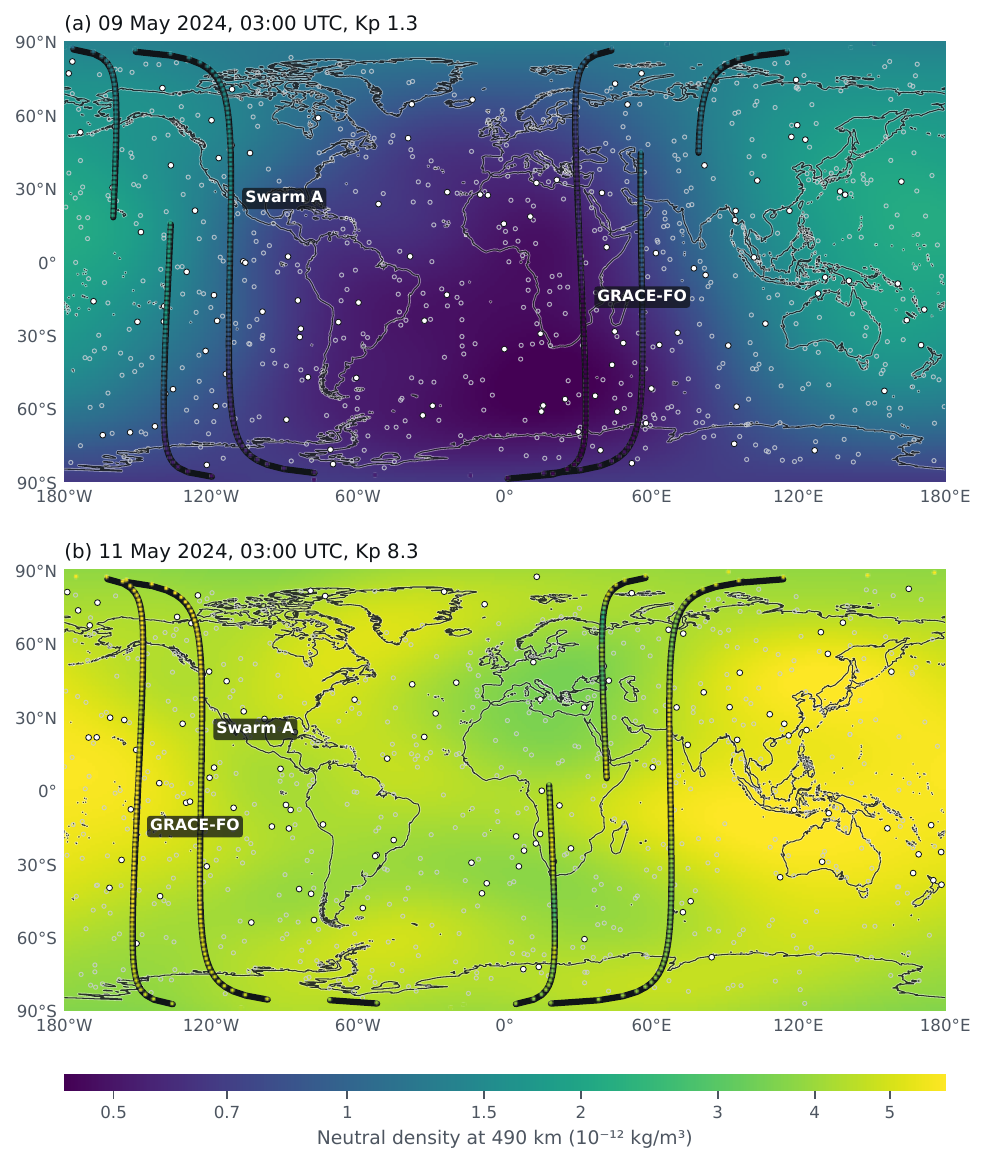}
\caption{\NDM\ density at 490~km before ((a), 9~May~2024 03:00~UT, \kp~1.3) and during ((b), 11~May~2024 03:00~UT, \kp~8.3) the May 2024 storm. Tracks: one orbit of GRACE-FO and Swarm~A, colored by the measured density on the same scale. Dots: the 668 TLE objects constraining \NDM\ on that day; filled dots are within $\pm$50~km of the slice, open dots at other altitudes (223--992~km).}
\label{fig:map}
\end{figure}

Objects and intervals are screened in three ways: objects are excluded if their median $B$ is not positive, if more than 20\% of their intervals imply a negative $B$, or if more than 5\% still show no decay at the maximum interval length; intervals are rejected if their implied $B$ differs from the object median by more than a factor of four; and residuals beyond five times $\sigma_i$ are clipped. Together with the Huber function, this screening also removes manoeuvres and mis-tagged objects. After screening, 586--802 passive objects per year are used (Figures~\ref{fig:map} and S4).

\textit{Calibration objects.} Five spheres with known area-to-mass ratio \cite<LCS-4 and Kosmos~1263, 1508, 2292 and 482;>{Mutschler2026} have $B$ fixed, so their decay directly measures the absolute density scale. Because these five objects contribute only a small fraction of the intervals, which number more than $10^5$ per year for the free objects, each of their intervals is given 30 times the normal weight ($\kappa_i=30$ in Equation~\ref{eq:loss}; $\kappa_i=1$ otherwise); without this the calibration would be lost among the free objects. These spheres fix the absolute scale where no accelerometer is available. Other geodetic spheres (Larets, Starlette, Stella) are left free and serve as a consistency check.

\textit{Accelerometer densities.} TU Delft densities from CHAMP, GRACE~A/B, GOCE, Swarm~A/B/C and GRACE-FO \cite{Siemes2016,vandenIJssel2020,Siemes2023} enter a Huber loss on $\ln(\rho_{\mathrm{obs}}/\rho_{\mathrm{MSIS}})-\delta$ at their published scale. This scale lies 4--9\% below the sphere scale, a difference consistent with differing drag-coefficient conventions; where accelerometers are available the product therefore follows the TU Delft scale, and where they are not it relaxes toward the sphere scale (Section~\ref{sec:sources}). Every year from 2003 to 2025 has at least three accelerometer satellites; 2026 is reconstructed from TLEs alone and is flagged as such in the product.

Figure~\ref{fig:map} illustrates the result: on 11~May~2024 the density at 490~km increases several-fold within a day, and the reconstruction agrees with the GRACE-FO and Swarm~A tracks while the $\sim$670 TLE objects that constrain it are spread over all longitudes, latitudes and altitudes.

\section{Validation Design}

Because \NDM\ assimilates every accelerometer satellite, its own fit to those satellites says little about accuracy. We therefore use two independent tests. (1)~\textit{Withheld satellites:} for each year, \NDM\ is retrained six times with identical configuration, each time leaving out one accelerometer group (CHAMP; GRACE~A/B; GOCE; Swarm~A/C; Swarm~B; GRACE-FO), and is scored against the withheld densities (60 satellite-years, 2003--2025). These retrainings are used only for evaluation and never enter the product. The same experiment is repeated with the TLEs removed (accelerometers only) and with the accelerometers removed (TLEs and spheres only) to isolate the contribution of each data source. (2)~\textit{Spire GNSS orbits:} precise orbits of 16--34 Spire CubeSats per year (November 2019--July 2025, 87,727 six-hour arcs, about 280--650~km) are never used in training. For each density model held fixed, a numerical propagator \cite<with GOCO06s gravity;>{Kvas2021} fits each satellite's ballistic coefficient on odd days; on even days the arcs are re-fitted with that coefficient fixed, and the remaining 6-h 3D position error measures how well the model reproduces the drag. Because $B$ is fitted, this test is insensitive to a constant density scale factor: it probes the spatial and temporal variability that matters for orbit prediction, not the absolute scale. It is important to emphasize that this is not a forecast test. The density used to propagate each arc is the reanalysis itself, built from data up to and beyond that time, and the test asks how consistent the reconstructed field is with orbits that played no part in building it.

For density accuracy we use a single metric, $\epsilon$, the root mean square of $\ln(\rho_{\mathrm{obs}}/\rho_{\mathrm{model}})$ over 94-min orbit averages. Because it is expressed in natural-log units, it is close to the rms fractional error when it is small, so that $\epsilon=0.10$ means the model is typically within about 10\% of the observations. We also split $\epsilon$ into a bias (the mean of the log ratio, a systematic scale offset) and a scatter (its standard deviation, the orbit-to-orbit variability).

Each withheld-satellite experiment, one accelerometer group left out for one calendar year, gives one test case, which we call a satellite-year. There are 60 such cases, and we report the median of $\epsilon$ over them. Their 95\% intervals are obtained by bootstrap, resampling whole weeks so that the correlation between consecutive days is preserved.

For Spire, the models are compared in pairs: for each satellite, we take the ratio of its position error with one model to its error with another, and report the median ratio over satellites, with 95\% intervals from resampling the satellites. When results are grouped by geomagnetic activity, a further complication arises: all satellites fly through the same storms, so satellites are not independent within a \kp\ bin. For these comparisons we therefore resample whole geomagnetic events, defined as runs of consecutive days in the same daily-maximum-\kp\ bin (5,000 resamples), and give no interval for bins with fewer than five events.

\NDM\ is compared with four baselines: \msis, NRLMSIS~2.1, the SET HASDM database \cite{Tobiska2021} extended to 2025, and, from July 2023, the operational WAM-IPE model of NOAA SWPC \cite{Fang2022}.

\section{Results}

\subsection{Density on Satellites Never Seen in Training}

\begin{figure}[ht]
\centering
\includegraphics[width=0.85\textwidth]{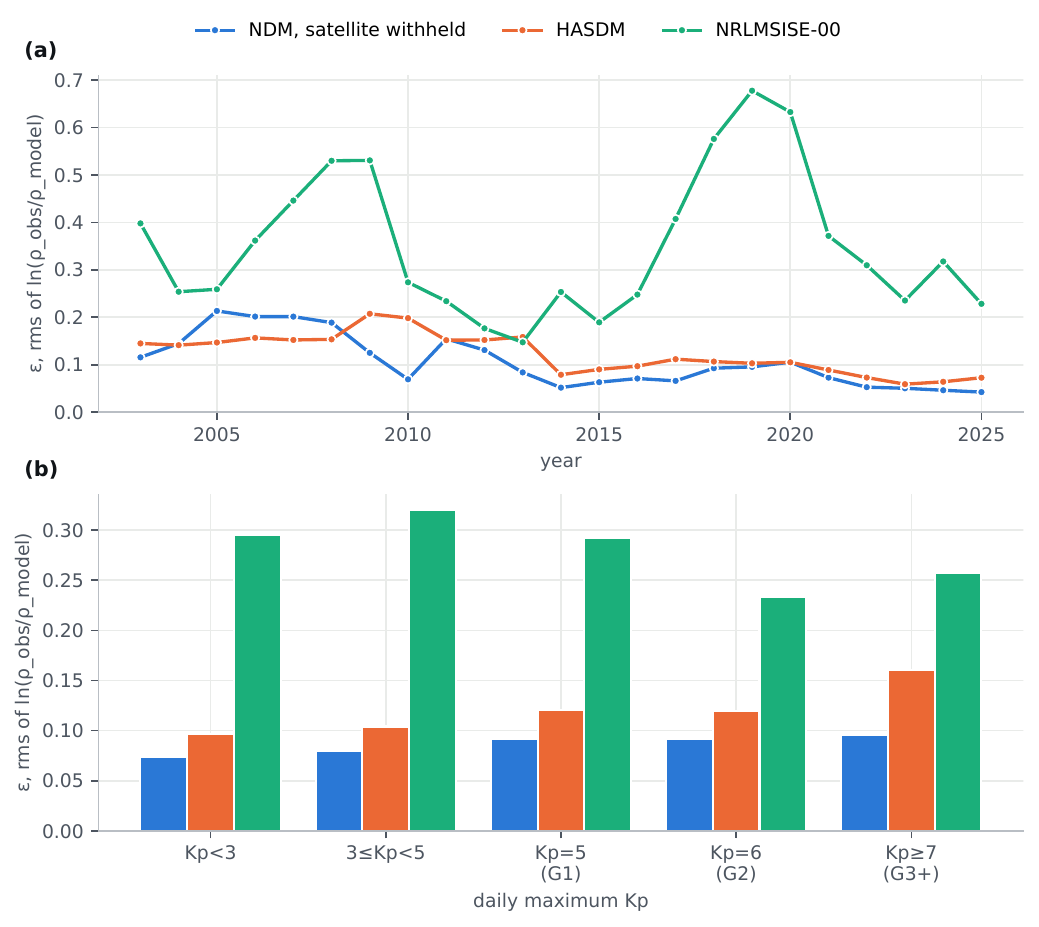}
\caption{Density error on accelerometer satellites withheld from training, $\epsilon$ = rms of $\ln(\rho_{\mathrm{obs}}/\rho_{\mathrm{model}})$ on 94-min orbit averages. (a) Median over the satellites withheld in each year, 2003--2025. (b) Median over the 60 satellite-years, by daily maximum \kp. Blue: \NDM\ retrained without the satellite; orange: HASDM; green: \msis.}
\label{fig:loo}
\end{figure}

Over the 60 withheld satellite-years the median error is $\epsilon=0.078$ for \NDM, 0.101 for HASDM and 0.310 for \msis\ (Figure~\ref{fig:loo}a, Table~\ref{tab:ablation}): 23\% below HASDM and a factor of four below the empirical model. \NDM\ outperforms HASDM in 45 of 60 cases, and in every year from 2009 onward except 2011 and 2020, where the two are tied. By daily maximum \kp\ (Figure~\ref{fig:loo}b) the median error of \NDM\ grows only from 0.074 to 0.095, while that of HASDM grows from 0.096 to 0.161.

The decomposition shows where this advantage comes from. The median bias is $-0.034$ for \NDM\ and $-0.074$ for HASDM, but the median scatter is 0.063 for \NDM\ and 0.052 for HASDM, which has the smaller scatter in 49 of 60 cases; on storm days (\kp~$\geq$~6) since 2013 the scatter is equal (0.074) while the biases are $+0.009$ and $-0.085$. \NDM's lower total error therefore comes mainly from its scale, learned from the other TU Delft satellites. The HASDM offset of about 7\% matches the 4--9\% offset between the TU Delft densities and the calibration spheres, so it may largely reflect a drag-coefficient convention rather than a model error; HASDM reproduces orbit-to-orbit variability slightly better, consistent with the Spire results below.

The reanalysis also quantifies the bias of \msis\ over two solar cycles. The global-mean correction at 400--500~km reaches $-50$ to $-57$\% during the 2008--2009 and 2019--2020 minima (Figure~S2), consistent with the record-low densities reported for the 2008 minimum \cite{Emmert2010}.

\subsection{What the Tracked Objects Contribute}
\label{sec:sources}

Table~\ref{tab:ablation} separates the data sources. From 2013 to 2025, adding the TLEs to the accelerometers lowers the median error by 20\% (0.083 to 0.066) and the 90th percentile from 0.146 to 0.099. More importantly, the TLEs remove the failures of accelerometer-only extrapolation, which reach $\epsilon=0.83$ (Swarm~A/C withheld in 2018) and 0.57 (2020); with TLEs the largest error after 2012 is 0.14. Trained on TLEs and spheres alone, with no accelerometer data, \NDM\ reaches a median of 0.15, half the error of \msis; this is the configuration of the 2026 product, for which we expect a median error of 0.11--0.14 (the TLE-only values in 2022--2025). On Spire orbits the TLE-only version reduces the \msis\ error by 20--44\% in 2022--2025 but by less than 10\% at the 2019--2021 solar minimum.

From 2003 to 2012 the picture is reversed: the TLEs raise the median error from 0.092 to 0.155, although they also remove the two worst accelerometer-only failures (0.55 and 0.35). In those years the accelerometers fly at only two or three altitudes, often 100--200~km apart, near solar minimum, where the TLE drag signal is weakest; the daily latent state then absorbs TLE noise, which increases day-to-day scatter at altitudes without accelerometer data. The error is smallest when a training satellite lies within 100~km in altitude of the evaluated orbit (Figure~S1).

Two further tests, both in the production configuration, address whether the observations, rather than the network, determine the spatial structure (Text~S1). In the first, a synthetic experiment, we prescribe a known density field made of separate components (a daily global level, an altitude gradient, a latitude pattern, a diurnal variation, small-scale structure, and 3-day and 6-h oscillations), generate the TLE observations it would produce with the real 2024 sampling, noise and drag operator, and train \NDM\ on them. For each component we report the gain, the ratio of recovered to true amplitude (1 means fully recovered, 0 not recovered at all), and, where useful, the correlation between recovered and true patterns. The TLEs alone recover the daily global level (gain 1.00, correlation $>$0.99), the altitude gradient (gain 0.98, correlation 0.97), the latitude structure (pattern correlation 0.81--0.87, but with an amplitude 18--29\% too large), the diurnal variation (gain 1.03--1.08, phase within 0.2~h below 700~km) and 3-day oscillations (gain 0.89--1.01). They do not recover small-scale latitude--local-time patterns or 6-h oscillations (gains near zero). Overall, the recovered spatial pattern correlates with the truth at 0.86--0.89. A simpler correction made of a daily scale factor with a daily altitude slope recovers the altitude gradient but no horizontal structure (grid error 0.10, against 0.065 for \NDM), and without the daily state the 3-day variability is recovered with a gain of only 0.28--0.35.

In the second test, on real data with GRACE-FO withheld (Table~S1), replacing the neural correction by a daily global scale factor raises $\epsilon$ from 0.050 to 0.100 in 2023 and from 0.046 to 0.074 in 2024, and on storm days it more than doubles it. Removing the daily state raises $\epsilon$ by only about 10\% (to 0.055 and 0.050), and on 2024 storm days not at all. Spatially resolved estimation thus provides most of the skill. The daily state is needed mainly for multi-day variability that the geomagnetic indices do not explain; structure at scales below the resolution above is inherited from \msis\ and the network.

\subsection{Five Major Storms of 2023--2024}

\begin{figure}[ht]
\centering
\includegraphics[width=0.85\textwidth]{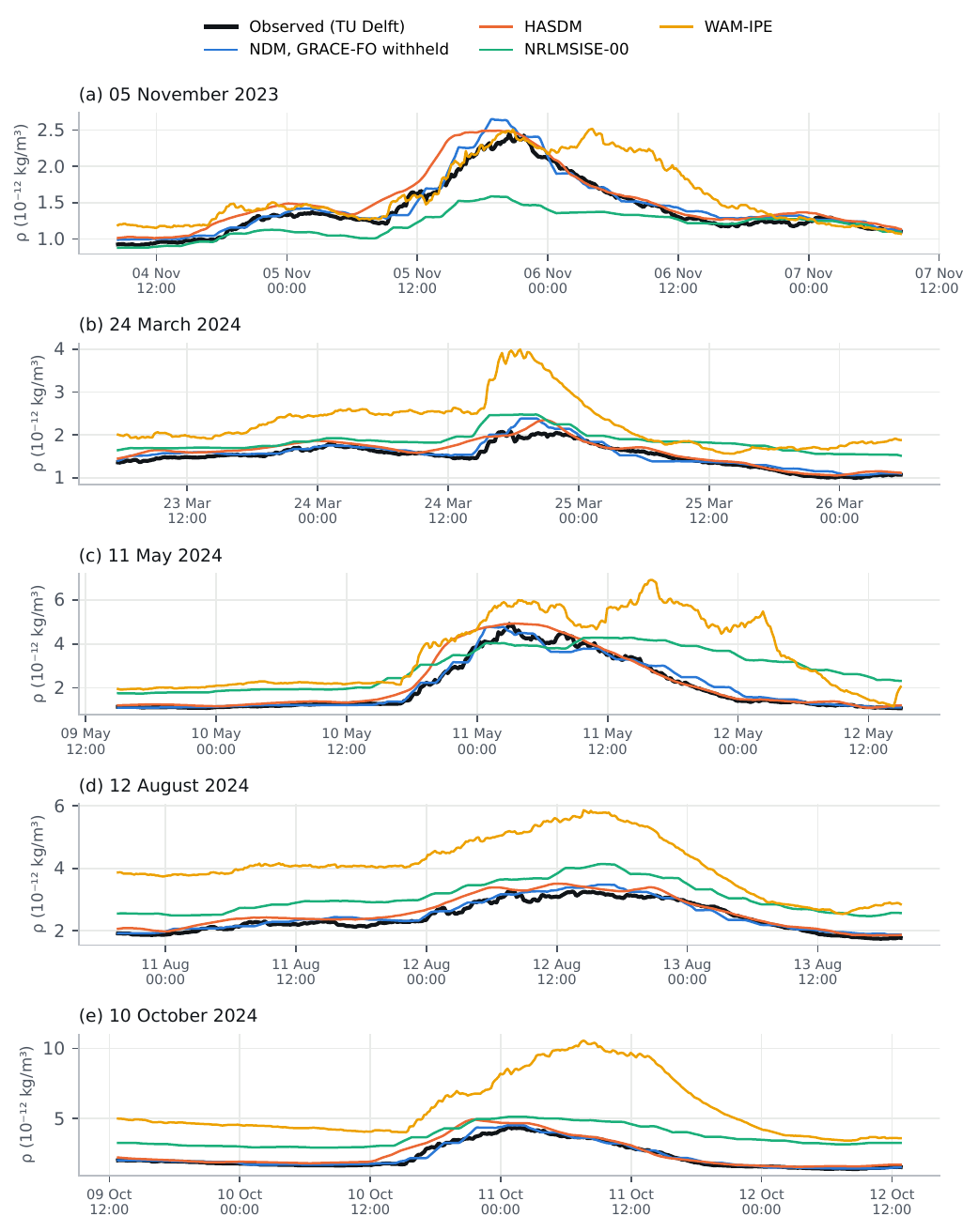}
\caption{Five major storms observed by GRACE-FO (94-min orbit averages, $\pm$1.5 days around each peak), with GRACE-FO withheld from \NDM. Black: observed (TU Delft). Seven-day errors $\epsilon$ for \NDM\ / HASDM / \msis\ / WAM-IPE: (a) 5~Nov~2023, 0.054 / 0.086 / 0.14 / 0.20; (b) 24~Mar~2024, 0.061 / 0.080 / 0.24 / 0.46; (c) 11~May~2024, 0.062 / 0.102 / 0.51 / 0.56; (d) 12~Aug~2024, 0.048 / 0.075 / 0.34 / 0.64; (e) 10~Oct~2024, 0.046 / 0.110 / 0.56 / 0.86.}
\label{fig:storms}
\end{figure}

Figure~\ref{fig:storms} shows five storms observed by GRACE-FO, with GRACE-FO withheld from \NDM\ (Swarm and all TLEs retained). All errors in this section are values of $\epsilon$ computed over a 7-day window around each storm, and biases are in the same natural-log units. \NDM\ reaches $\epsilon=0.046$--0.062, against 0.075--0.110 for HASDM, 0.14--0.56 for \msis\ and 0.20--0.86 for WAM-IPE. For the May 2024 (Gannon) storm, $\epsilon$ is 0.51 for \msis\ and 0.56 for WAM-IPE, while \NDM\ follows the observed rise and recovery with $\epsilon=0.062$. \NDM\ is nearly unbiased (bias $-0.02$ to $-0.04$), whereas HASDM has a bias of $-0.07$ to $-0.09$, meaning that it overestimates the observed density by 7--9\%, again comparable to the offset between the TU Delft and sphere scales. The \NDM\ main phase is often slightly late or smoothed, a consequence of the daily state and the 1--3-day TLE intervals (Section~\ref{sec:discussion}).

\subsection{Independent Test on Spire GNSS Orbits}

\begin{figure}[ht]
\centering
\includegraphics[width=0.9\textwidth]{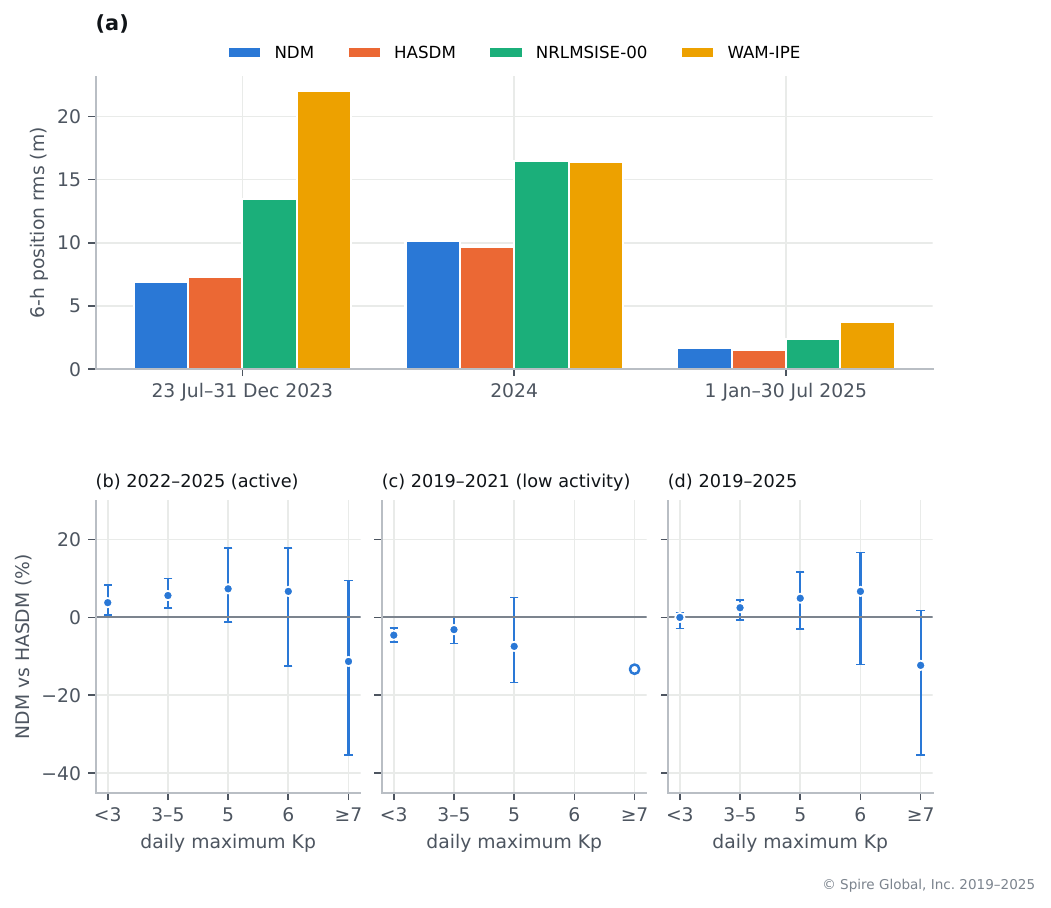}
\caption{Independent test on Spire GNSS orbits (6-h arcs, ballistic coefficient fixed from other days). (a) Median 6-h position error in the WAM-IPE period (23~Jul~2023--30~Jul~2025), same arcs for all models. (b--d) Paired median change in position error of \NDM\ relative to HASDM, by daily maximum \kp, for (b) 2022--2025, (c) 2019--2021 and (d) 2019--2025; negative values mean \NDM\ is better. Error bars are 95\% intervals from resampling whole geomagnetic events; open symbols mark bins with fewer than five events, for which no interval is given. \copyright~Spire Global, Inc. 2019--2025.}
\label{fig:spire}
\end{figure}

On Spire orbits \NDM\ reduces the 6-h position error relative to \msis\ in every year from 2019 to 2025 (Table~\ref{tab:spire}). The improvement is statistically significant in every year: the 95\% interval of the median ratio of the two errors, taken satellite by satellite, lies entirely below one. Relative to HASDM, the result depends on solar activity and altitude: \NDM\ is better by 4--5\% during the 2019--2021 solar minimum, tied in 2023, and behind by 5--16\% in 2022, 2024 and 2025. The HASDM advantage lies mainly at 550--650~km and 400--450~km, outside the altitude band of the training accelerometers. In the WAM-IPE period (Figure~\ref{fig:spire}a) \NDM\ reduces the error of WAM-IPE by 38--69\%; in 2024 WAM-IPE is no better than \msis. The ballistic coefficients fitted with \NDM\ or HASDM scatter by 6--8\% from arc to arc in 2022--2025, against 15--17\% with \msis, a direct measure of improved orbit predictability. Since $B$ is fitted, these results concern the variability of the field, not its absolute scale.

By geomagnetic activity (Figures~\ref{fig:spire}b--\ref{fig:spire}d), \NDM\ and HASDM agree to within a few percent for \kp~$<$~7. In 2022--2025, HASDM, which assimilates radar tracking, has errors 4--7\% lower, a difference that is significant for \kp~$<$~5. On days with \kp~$\geq$~7, the median error of \NDM\ is 12\% lower than that of HASDM over 2019--2025, and its error relative to \msis\ is $-59$\% against $-53$\% for HASDM (Figure~S3). This category, however, contains only 13 storm events with test arcs, two of them in 2019--2021 (Table~S2). Resampling whole events gives a 95\% interval for the difference that ranges from \NDM\ being 35\% better to 2\% worse: \NDM\ is better in 8 of the 13 events, but HASDM is better in the two largest (May and October 2024). Because this interval includes zero, the data are consistent with no real difference between the two models in severe storms; the 12\% advantage of \NDM\ is the best single estimate, but with only 13 storms it cannot be considered established.

\section{Discussion and Conclusions}
\label{sec:discussion}

\NDM\ shows that the orbital decay of a few hundred passive objects per year, combined with a few known-shape spheres and the accelerometer record, constrains the global 3D thermospheric density with skill comparable to the operational product that assimilates non-public radar data: lower total error on withheld accelerometers, mainly through a smaller bias, and similar performance on independent orbits.

The key methodological step is to fit the density model directly to the observed orbital decay. In the conventional use of TLEs \cite{Picone2005,Emmert2008,Doornbos2008}, the decay of each object over each interval is first converted into a single density value: an average weighted along the orbit, which depends on an assumed ballistic coefficient and is assigned to one location, usually near perigee, through an assumed altitude profile. Errors in the assumed coefficient and profile, and the loss of information about where along the orbit the drag occurred, then enter every subsequent step. In \NDM\ no such intermediate densities exist. The fully differentiable chain from the network parameters through Equation~\ref{eq:chain} to the observed changes in mean motion compares each observation with exactly the quantity it measures, the drag integrated along the actual trajectory. This allows the density field, the daily state and $\sim$700 ballistic coefficients to be estimated jointly and consistently from heterogeneous data.

The second choice that makes \NDM\ possible is that the neural correction is spatially resolved. In contrast to the few global drivers estimated by most assimilation schemes \cite{Brandt2020,Sutton2021,Forootan2023,Mutschler2026}, \NDM\ learns latitude, local-time and altitude structure from objects in many orbital planes, which a daily global scale factor cannot reproduce (Section~\ref{sec:sources}).

What is observation-constrained should be distinguished from what is model-driven. The daily level, the altitude gradient, the latitude pattern (although not its exact amplitude) and the diurnal variation below about 700~km, at time scales of a day or longer, are constrained by the data. Sub-daily storm evolution and small-scale structure come from the geomagnetic drivers, the network's learned response to them and \msis. Storm main phases are therefore under-estimated and lagged by hours \cite<cf.>{Constant2026}; shorter TLE intervals did not remove this lag, and resolving it will require precise orbits, for example through a full differentiable orbit propagator applied to GNSS-tracked satellites.

The current version has further limitations. (1) In 2003--2012 the TLEs increase the median error relative to accelerometers alone, and away from accelerometer altitudes near solar minimum the TLE noise increases day-to-day scatter. (2) The absolute scale is uncertain by about $\pm$10\%, from the area-to-mass ratios of the calibration spheres and the modeled drag coefficient, and the TU Delft and sphere scales differ by 4--9\%. (3) The yearly networks produce seams at year boundaries with a near-zero mean jump and an rms jump of 0.02--0.05, about twice the seed spread; the archived product will include the $\pm$15-day overlaps so that users can blend across boundaries. (4) The 2026 field is constrained by TLEs alone. Without accelerometers the scale relaxes toward the sphere scale: in 2022--2025 the TLE-only fields lie 0.02--0.12 (ln units, increasing with altitude from 300 to 700~km) above the product, so 2026 is expected to lie above the TU Delft level by about this amount, with a median error of 0.11--0.14 based on the retrospective TLE-only validation. (5) Uncertainty quantification is not yet included. The seed spread (0.006--0.016) is much smaller than the validation errors and is not a usable uncertainty. Until a full error model is available, the validation errors by altitude gap, solar activity and \kp\ (Figures~\ref{fig:loo} and S1) provide an empirical guide: typically 0.04--0.10 within 100~km of an accelerometer satellite after 2012, and up to 0.15--0.30 for upward extrapolation at solar minimum. (6) In the synthetic test the amplitude of the latitude structure is overestimated by 18--29\% (13--18\% without noise), so the overshoot comes partly from the sampling and not only from the noise.

In conclusion, the thermosphere has so far lacked what weather forecasting gained from reanalyses: a long, continuous and observation-constrained record of the 3D state that anyone can reproduce. \NDM\ provides such a record for 2003--2026. It combines the decay of tracked debris and rocket bodies with a few calibration spheres and the accelerometer missions, and it reaches an accuracy on unseen satellites comparable to that of HASDM and four times better than that of \msis. Because it relies only on public observations, it can be extended in near-real time as new TLEs and new missions become available. We expect it to be useful in three ways: as a reference for orbit determination and collision avoidance, as a benchmark for physics-based and empirical models, and as a training set for the next generation of machine-learning forecasts of thermospheric density \cite<e.g.,>{Licata2022,Wang2026}.

\section*{Open Research}

TLEs are available from Space-Track.org (\url{https://www.space-track.org}, free registration). Accelerometer and GNSS-derived densities of CHAMP, GRACE, GOCE, Swarm and GRACE-FO are available from TU Delft (\url{http://thermosphere.tudelft.nl}). Geomagnetic and solar indices are from GFZ Potsdam \cite{Matzka2021}. Spire GNSS orbits were obtained through the NASA Commercial Satellite Data Acquisition (CSDA) Program and are available to eligible users from NASA CSDA. SET HASDM densities are available for scientific use on request from Space Environment Technologies \cite{Tobiska2021}. WAM-IPE output is distributed by NOAA SWPC. \msis\ and NRLMSIS~2.1 were evaluated with the open-source \texttt{pymsis} package. The \NDM\ reanalysis (daily 3D grids, 2003--2026) and the code used to produce it will be deposited at [Zenodo DOI to be added upon acceptance] under an open licence, together with the configuration files and scripts needed to reproduce the withheld-satellite and Spire experiments. The archived fields include the $\pm$15-day overlaps between yearly networks and a flag marking the TLE-only period (2026).

\acknowledgments
This work utilized data made available through the NASA Commercial Satellite Data Acquisition (CSDA) Program; Spire data \copyright~Spire Global, Inc. 2019--2025. The SET HASDM density data are provided for scientific use courtesy of Space Environment Technologies. We thank TU Delft for the accelerometer and POD densities, NOAA SWPC for WAM-IPE output, GFZ Potsdam for the geomagnetic and solar indices, the 18th Space Defense Squadron and Space-Track.org for TLEs, and Natural Earth for coastlines.

\newpage

\begin{table}[ht]
\caption{Spire 6-h 3D Position Error (m) With Fixed Ballistic Coefficient$^{a}$}
\centering
\begin{tabular}{l r r r r r r}
\hline
Year & Arcs & \msis & NRLMSIS 2.1 & HASDM & \NDM & \NDM\ vs HASDM \\
\hline
2019 (Nov--Dec) & 3,242  & 0.41  & 0.40  & 0.41          & \textbf{0.37}  & $-5$\% (n.s.) \\
2020            & 12,530 & 0.61  & 0.58  & 0.52          & \textbf{0.48}  & $-4.5$\% [$-6$, $-3$] \\
2021 (to Nov)   & 15,770 & 0.66  & 0.64  & 0.54          & \textbf{0.51}  & $-4$\% [$-7$, $-2$] \\
2022            & 26,289 & 2.62  & 2.58  & \textbf{1.24} & 1.29           & $+5.5$\% [$+2$, $+8$] \\
2023            & 21,765 & 11.69 & 11.17 & \textbf{4.89} & 5.06           & $+0.1$\% [$-2$, $+6$] \\
2024            & 3,393  & 16.45 & 20.01 & \textbf{9.68} & 10.10          & $+16$\% [$+7$, $+27$] \\
2025 (to Jul)   & 4,738  & 2.36  & 2.40  & \textbf{1.54} & 1.62           & $+5$\% [$+2$, $+15$] \\
\hline
\multicolumn{7}{p{0.95\textwidth}}{\footnotesize $^{a}$Median over satellites of the test-day (even-day) 6-h position rms; ballistic coefficient from odd days. Last column: for each satellite, the ratio of its \NDM\ error to its HASDM error; the median ratio minus one, with a 95\% interval from resampling satellites (negative = \NDM\ better). The same comparison against \msis\ favors \NDM\ significantly in every year.}
\end{tabular}
\label{tab:spire}
\end{table}

\begin{table}[ht]
\caption{Density Error on Withheld Satellites by Data Source$^{a}$}
\centering
\begin{tabular}{l c c c}
\hline
Model & 2003--2012 (22) & 2013--2025 (38) & All (60) \\
\hline
\NDM\ (TLEs + spheres + accelerometers) & 0.155 / 0.223 & \textbf{0.066} / \textbf{0.099} & \textbf{0.078} / \textbf{0.180} \\
\NDM, accelerometers only               & \textbf{0.092} / 0.331 & 0.083 / 0.146 & 0.084 / 0.228 \\
\NDM, TLEs + spheres only               & 0.178 / 0.260 & 0.144 / 0.212 & 0.151 / 0.236 \\
HASDM                                   & 0.176 / \textbf{0.211} & 0.088 / 0.121 & 0.101 / 0.206 \\
\msis                                   & 0.320 / 0.530 & 0.310 / 0.642 & 0.310 / 0.602 \\
\hline
\multicolumn{4}{p{0.9\textwidth}}{\footnotesize $^{a}$Median / 90th percentile over satellite-years (number in parentheses) of $\epsilon$, the rms of $\ln(\rho_{\mathrm{obs}}/\rho_{\mathrm{model}})$ on 94-min orbit averages. All \NDM\ versions share the production configuration; the evaluated satellite is never used in training. Largest errors: 0.29 (\NDM), 0.83 (accelerometers only), 0.33 (TLEs + spheres only), 0.23 (HASDM).}
\end{tabular}
\label{tab:ablation}
\end{table}

\newpage
\bibliography{references}

\end{document}


\title{Supporting Information for ``The Neural Density Model: An Open 3D Thermospheric Density Reanalysis for 2003--2026 From Tracked Space Debris''}
   
\authors{Enrico Camporeale\affil{1,2}}
\affiliation{1}{School of Physical and Chemical Sciences, Queen Mary University of London, London, UK}
\affiliation{2}{Space Weather Technology, Research and Education Center (SWx TREC), University of Colorado Boulder, Boulder, CO, USA}

\noindent\textbf{Contents of this file}
\begin{enumerate}
\item Text S1
\item Tables S1 and S2
\item Figures S1 to S4
\end{enumerate}

\noindent\textbf{Introduction}
Text S1 and Table S1 describe two tests, in the production configuration, of which parts of the density field are constrained by the TLE observations. Table S2 lists the severe storms in the Spire validation period. Figures S1--S4 give the breakdown of the density and Spire validation by altitude, solar activity, local time and geomagnetic activity, the global-mean correction of NRLMSISE-00 by the NDM reanalysis over 2003--2026, and the data used in each period. All density errors are the rms of $\ln(\rho_{\mathrm{obs}}/\rho_{\mathrm{model}})$ on 94-min orbit averages, computed with NDM retrained without the evaluated satellite.

\renewcommand{\thetable}{S\arabic{table}}

\noindent\textbf{Text S1. Identifiability tests.}
Both tests use the production configuration. This means the production flags and regularization of the year, all TLE objects, and the five calibration objects at their fixed area-to-mass ratios. The drag coefficient is computed per sample with the same operator as in training.

\textit{Synthetic recovery.} Synthetic mean-motion changes were generated on the real screened TLE intervals of 2024 from a known correction $\delta_{\mathrm{true}}$. The ballistic coefficients were taken from the production run, and the per-object noise from that run's TLE residuals. The truth contained a daily global level (random walk, step 0.04), an altitude gradient (0.15 per 200~km, modulated by $\pm$50\% over 27 days), a latitude $P_2$ term (0.10), a diurnal term ($\cos(\mathrm{LST}-14\,\mathrm{h})\cos\phi$, 0.10), a small-scale term ($\sin 3\,\mathrm{LST}\,\sin 2\phi$, 0.05), a 3-day oscillation (0.06) and a 6-h oscillation (0.05). \NDM\ was trained on TLEs alone, with realistic noise and, for comparison, without noise. Over 300--800~km it recovered:
\begin{itemize}
\item the daily level with gain 1.00 (correlation $\geq$0.997);
\item the altitude gradient with gain 0.98 (slope correlation 0.97);
\item the latitude structure with pattern correlation 0.81--0.87, but an amplitude 18--29\% too large (13--18\% without noise);
\item the diurnal term with gain 1.03--1.08, pattern correlation 0.75--0.77, and a phase error $\leq$0.2~h below 700~km (0.6~h at 800~km);
\item the 3-day oscillation with gain 0.89--1.01.
\end{itemize}
The small-scale and 6-h terms were not recovered (gain $\approx$0). With the global mean removed at each time step, the recovered spatial pattern correlates with the truth at 0.86--0.89 and explains 63--68\% of its variance; the unrecoverable small-scale term limits this value. Three reduced corrections were trained on the same synthetic data:
\begin{itemize}
\item A daily scale factor with a daily altitude slope, $\delta=c_d+s_d\,(h-500\,\mathrm{km})/200\,\mathrm{km}$, recovers the altitude gradient (gain 1.12) but no horizontal structure; its grid error is 0.10, against 0.065 for \NDM.
\item A daily scale factor alone recovers neither.
\item Without the daily state, the spatial pattern is recovered but the 3-day oscillation is not (gain 0.28--0.35).
\end{itemize}

\textit{Reduced corrections on real data.} Table~S1 compares the full correction with the three reduced forms, using the production withheld-satellite protocol: GRACE-FO withheld for the whole year, with Swarm~A/B/C and all TLEs kept. Because the altitude slope is defined about 500~km and GRACE-FO flies near 490~km, the scale-plus-slope variant cannot differ from the scale-only variant on this test; it is evaluated by the synthetic experiment above.

\begin{table}[h]
\caption{Error $\epsilon$ on Withheld GRACE-FO for Reduced Forms of the Correction (Production Configuration)$^{a}$}
\centering
\begin{tabular}{l c c c c}
\hline
Correction & 2023 & 2024 & 2023 storm days & 2024 storm days \\
\hline
Full: $c_d + f_\theta(\mathbf{x},\mathbf{z}_d)$ & \textbf{0.050} & \textbf{0.046} & \textbf{0.066} & 0.061 \\
Daily scale factor: $c_d$ & 0.100 & 0.074 & 0.157 & 0.143 \\
Daily scale and altitude slope & 0.100 & 0.074 & 0.158 & 0.143 \\
No daily state: $f_\theta(\mathbf{x})$ & 0.055 & 0.050 & 0.070 & \textbf{0.056} \\
HASDM & 0.075 & 0.082 & 0.106 & 0.111 \\
NRLMSISE-00 & 0.244 & 0.340 & 0.167 & 0.388 \\
\hline
\multicolumn{5}{p{0.85\textwidth}}{\footnotesize $^{a}$Orbit-averaged rms of $\ln(\rho_{\mathrm{obs}}/\rho_{\mathrm{model}})$ over the calendar year; storm days have daily maximum $K_p\geq6$.}
\end{tabular}
\end{table}

\begin{table}[h]
\caption{Storms With Daily Maximum $K_p \geq 7-$ During the Spire Coverage (November 2019--July 2025)$^{a}$}
\centering
\begin{tabular}{l c c | l c c}
\hline
Start & Days & Max $K_p$ & Start & Days & Max $K_p$ \\
\hline
2021-05-12 & 1 & 7.0 & 2024-05-10 & 3 & 9.0 \\
2021-11-04 & 1 & 7.7 & 2024-06-28 & 1 & 7.7 \\
2022-04-10 & 1 & 6.7 & 2024-08-04 & 1 & 7.0 \\
2022-08-17 & 1 & 6.7 & 2024-08-12 & 1 & 8.0 \\
2023-02-27 & 1 & 6.7 & 2024-09-12 & 1 & 7.0 \\
2023-03-23 & 2 & 8.0 & 2024-09-17 & 1 & 7.3 \\
2023-04-23 & 2 & 8.3 & 2024-10-08 & 3 & 8.7 \\
2023-11-05 & 1 & 7.3 & 2025-01-01 & 1 & 8.0 \\
2023-12-01 & 1 & 6.7 & 2025-04-16 & 1 & 7.7 \\
2024-04-19 & 1 & 7.0 & 2025-06-01 & 3 & 7.7 \\
2024-05-02 & 1 & 6.7 & 2025-06-13 & 1 & 6.7 \\
\hline
\multicolumn{6}{p{0.8\textwidth}}{\footnotesize $^{a}$GFZ $K_p$; days within two days of each other are merged into one storm; only months with Spire data are counted. Sixteen storms reach $K_p \geq 7.0$ and six more reach $7-$ (6.7). Thirteen of these storms contribute test arcs to the $K_p \geq 7$ category of Figure~4 (11 in 2022--2025, 2 in 2019--2021).}
\end{tabular}
\end{table}

\clearpage
\begin{figure}
\centering
\includegraphics[width=\textwidth]{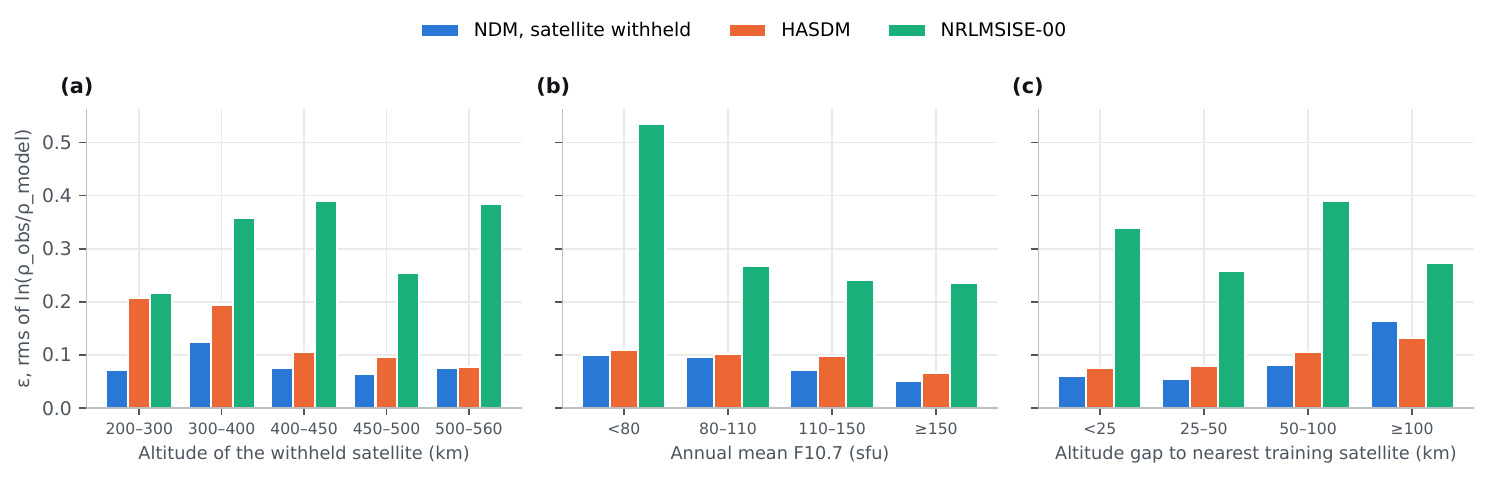}
\caption{Density error on withheld accelerometer satellites, by altitude of the withheld satellite, annual mean F10.7, and altitude gap to the nearest accelerometer satellite remaining in training. NDM is outperformed by HASDM only when the gap exceeds 100~km (mostly GRACE withheld in 2004--2012). HASDM data courtesy of Space Environment Technologies.}
\label{fig:s1}
\end{figure}

\begin{figure}
\centering
\includegraphics[width=\textwidth]{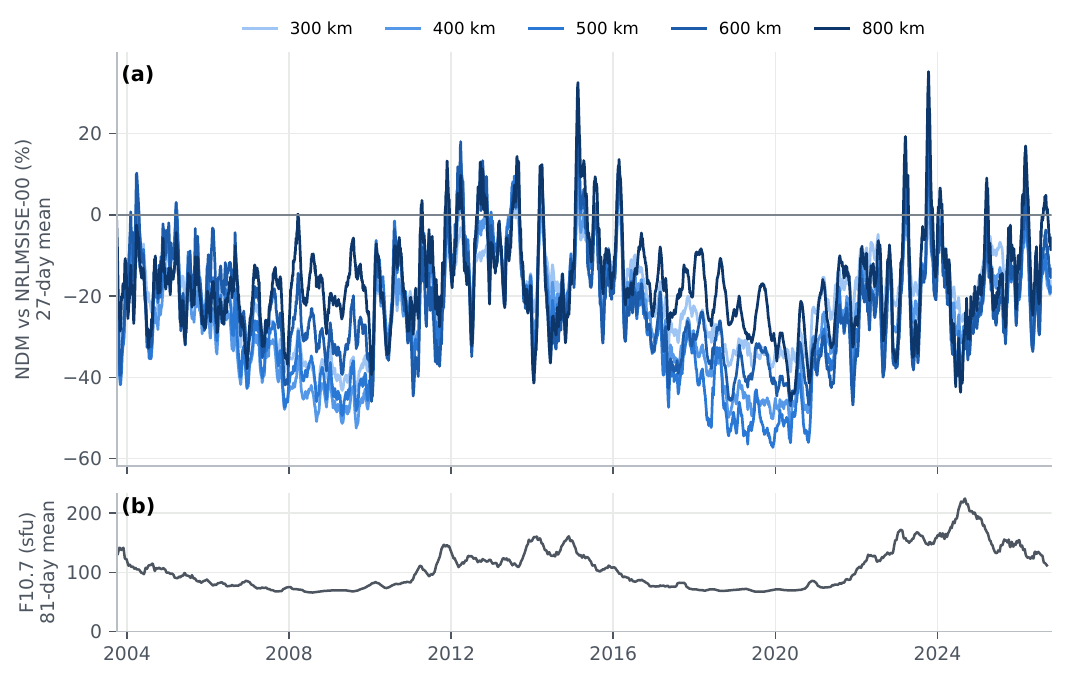}
\caption{Global-mean correction of NRLMSISE-00 by the NDM reanalysis at 300--800~km (27-day running mean, top) and 81-day mean F10.7 (bottom), October 2003--October 2026. The reanalysis lowers NRLMSISE-00 by 50--57\% at 400--500~km during the 2008--2009 and 2019--2020 solar minima.}
\label{fig:s2}
\end{figure}

\begin{figure}
\centering
\includegraphics[width=\textwidth]{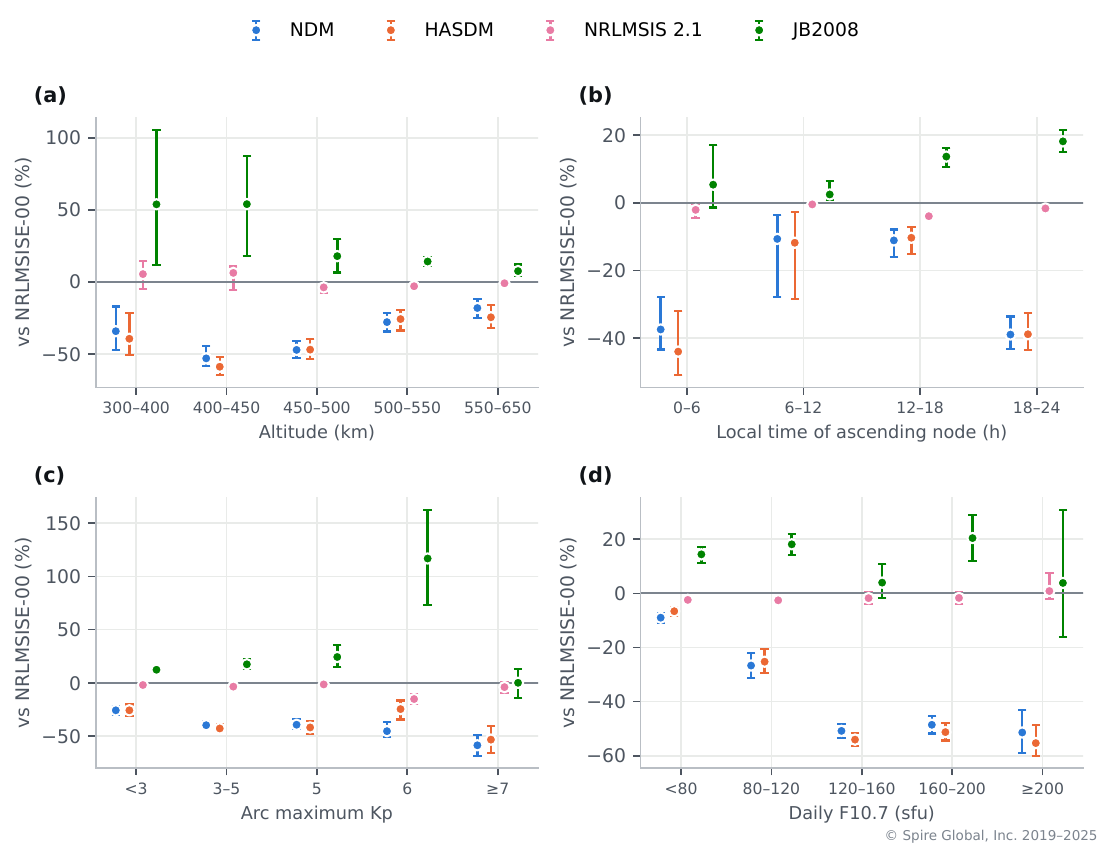}
\caption{Spire 6-h position error relative to NRLMSISE-00 (median change, 95\% bootstrap interval; negative = better), by altitude, local time of the ascending node, arc maximum $K_p$ and daily F10.7, for NDM, HASDM, NRLMSIS~2.1 and JB2008. JB2008 was run as configured in the \texttt{pyatmos} package; its setup has not been independently verified. \copyright~Spire Global, Inc. 2019--2025; data obtained through NASA CSDA.}
\label{fig:s3}
\end{figure}

\begin{figure}
\centering
\includegraphics[width=\textwidth]{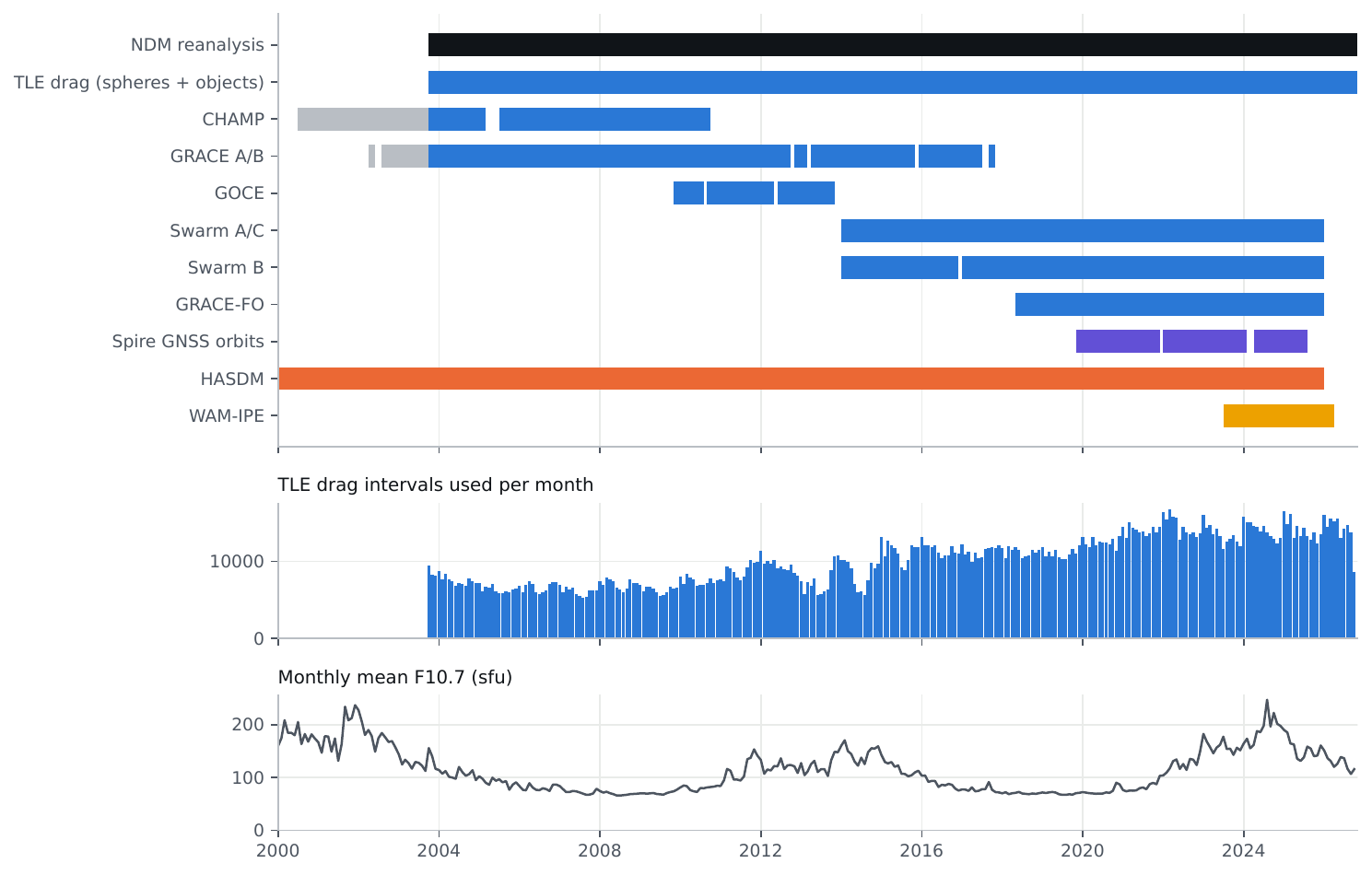}
\caption{Data used by NDM (blue), available but not used (grey), used for validation only (violet), and comparison models (orange, yellow), 2000--2026; TLE drag intervals used per month; monthly mean F10.7.}
\label{fig:s4}
\end{figure}

\clearpage